\documentclass[12pt]{article}

\usepackage{amsfonts,amssymb, amsmath,mathrsfs}

\usepackage[english]{babel}

\DeclareMathAlphabet{\mathpzc}{OT1}{pzc}{m}{it}

\def\on#1#2{\mathop{\vbox{\ialign{##\crcr\noalign{\kern2pt}
$\scriptstyle{#2}$\crcr\noalign{\kern2pt\nointerlineskip}
\kern-2pt$\hfil\displaystyle{#1}\hfil$\crcr}}}\limits}

\def\nn{ \nonumber }
\def\bq{ \begin{equation} }
\def\eq{ \end{equation} }
\def\ben{ \begin{eqnarray} }
\def\en{ \end{eqnarray} }

\def\e{{\rm e}}

\newtheorem{prop}{Proposition}

\newtheorem{exa}{Example}
\newenvironment{exam}{\begin{exa} \rm }{\end{exa}}

\newtheorem{re}{Remark}

\begin{document}


\title{Addition theorems and  the Drach superintegrable systems}

\author{A V Tsiganov \\
\it\small
St.Petersburg State University, St.Petersburg, Russia\\
\it\small e--mail: tsiganov@mph.phys.spbu.ru}
\date{}
 \maketitle

\begin{abstract}
 We propose new construction of the polynomial integrals of motion related to the addition theorems. As an example we reconstruct Drach systems and get some new two-dimensional superintegrable St\"ackel systems with third, fifth and seventh order integrals of motion.
\end{abstract}

\section{Introduction}
The Liouville classical theorem on completely integrable Hamiltonian systems implies that almost all points of the manifold $M$ are covered by a system of open toroidal domains with the action-angle coordinates $I=(I_1,\ldots,I_k)$ and $\omega=(\omega_1,\ldots,\omega_n)$:
\bq\label{aa-br}
\{I_j,I_k\}=\{\omega_i,\omega_k\}=0,\qquad \{I_j,\omega_k\}=\delta_{ij}.
\eq
The independent integrals of motion $H_1,\ldots,H_n$ are functions of the independent action variables $I_1,\ldots,I_n$ and the corresponding Jacobian does not equal to zero
\bq\label{h-mat}
\det\mathbf J\neq 0,\qquad\mbox{where} \qquad \mathbf J_{ij}= \dfrac{\partial H_i(I_1,\ldots,I_n)}{\partial I_j}\,.
\eq
Let us introduce $n$ functions
\bq\label{phi_k}
\phi_j=\sum_{k} \left(\mathbf J^{-1}\right)_{kj}\,\omega_k,\eq
such that
\bq\label{tr-alg}
\{H_i,\phi_j\}=\sum_{k=1}^{n} \mathbf J_{ik}\left(\mathbf J^{-1}\right)_{kj} =\delta_{ij}.
\eq
If Hamiltonian $H=H_1$ then the ($n-1$) functions $\phi_2,\ldots,\phi_n$ are integrals of motion
\[
\dfrac{d\phi_j}{dt}=\{H_1,\phi_j\}=0,\qquad j=2,\ldots,n,
\]
which are functionally independent on $n$ functions $H_1(I),\ldots,H_n(I)$.

So, in classical mechanics any completely integrable system is superintegrable system in a neighborhood of any regular point of $M$ \cite{ts07f}. It means that the Hamiltonian $H=H_1$ has $2(n-1)$ integrals of motion $H_2,\ldots,H_n$ and $\phi_2,\ldots,\phi_n$ on any open toroidal domain.

If the action-angle variables are global variables on the whole phase space $M$ and, therefore, we have superintegrable systems on $M$. For instance, the global action-angle variables for the open and periodic Toda lattices are discussed in \cite{hk08}.

However, in generic case the angle variables $\omega_k$ are multi-valued functions on the whole phase space $M$. If we have $k$ additional single-valued algebraic integrals of motion $K$ the trajectories are closed (more generally, they are constrained to an $n-k$ dimensional manifold in phase space).

Any additional integral is a function on the action-angle variables. Since we have to understand how to get single-valued additional integrals of motion from the multi-valued action-angle variables. In this paper we discuss a possibility to get polynomial integrals of motion from the multi-valued angle variables by using simplest addition theorem.

\section{The St\"{a}ckel systems.}
\setcounter{equation}{0}

The system associated with the name of St\"{a}ckel \cite{st95,ts99} is a holonomic system on the phase space $M=\mathbb R^{2n}$, with the canonical variables $q=(q_1,\ldots,q_n)$ and $p=(p_1,\ldots,p_n)$:
\bq \Omega=\sum_{j=1}^n dp_j\wedge dq_j\,,\qquad
\{p_j,q_k\}=\delta_{jk}\,.\label{stw}
\eq
The nondegenerate $n\times n$ St\"{a}ckel matrix $S$, whose $j$ column depends on the coordinate $q_j$ only, defines $n$ functionally independent integrals of motion
\bq
H_k=\sum_{j=1}^n ( S^{-1})_{jk}\Bigl(p_j^2+U_j(q_j)\Bigr)\,.
\label{fint}
\eq
From this definition one immediately
gets the separated relations
\bq
p_j^2=
\sum_{k=1}^n H_k S_{kj}-U_j(q_j)\,\label{stc}
\eq
and the angle variables
\[
\omega_i=\sum_{j=1}\int\,\dfrac{S_{ij}\,\mathrm dq_j}{p_j}=
\sum_{j=1}\int\,\frac{S_{ij}\,\mathrm dq_j}{\sqrt{\sum_{k=1}^n H_k S_{kj}-U_j(q_j)  } }\,.
\]
It allows reducing  solution of the equations of motion to a problem in algebraic geometry \cite{ts99}. Namely, let us suppose that there are functions $\mu_j$ and $\lambda_j$ on the canonical separated variables
\bq\label{fr-2}
\mu_j=u_j(q_j)p_j,\qquad \lambda_j=v_j(q_j)\,,\qquad \{q_i,p_j\}=\delta_{ij},
\eq
which allows us to rewrite separated equations (\ref{stc}) as equations defining the hyperelliptic curves
\bq\label{sthc}
\mathcal C_j:\quad \mu_j^2=P_j(\lambda_j)\equiv u_j^2(\lambda_j)\left(\sum_{k=1}^n H_k S_{kj}(\lambda_j)-U_j(\lambda_j)\right),
\eq
where $P_j(\lambda_j)$ are polynomials on $\lambda_j$. In this case
the action variables $I_k=H_k$ (\ref{fint}) have the canonical Poisson brackets (\ref{aa-br}) with the  angle variables
\bq\label{w-st}
\omega_i=\sum_{j=1}^n\int_{A_j} \dfrac{S_{ij}(\lambda_j)}{\sqrt{P_j(\lambda_j)\,}}\,\mathrm d\lambda_j=\sum_{j=1}^n \vartheta_{ij}(p_j,q_j),
\eq
which are the sums of integrals $\vartheta_{ij}$ of the first kind Abelian differentials on the hyperelliptic curves $\mathcal C_j$ (\ref{sthc})  \cite{ts99,ts07f}, i.e. they are sums of the
multi-valued functions on the whole phase space.

\subsection{Addition theorems and algebraically superintegrable systems}

In generic case the action variables (\ref{w-st}) are the sum of the multi-valued functions $\vartheta_{ij}$. However, if we are able to apply some addition theorem to the calculation of $\omega_i$ (\ref{w-st})
\bq\label{add-th}
\omega_i=\sum_{j=1}^n \vartheta_{ij}(p_j,q_j)=\Theta_i \bigl(K_i\bigr)+const,
\eq
where $\Theta_i$ is a multi-valued function on the algebraic argument $K_i(p,q)$  then one will get algebraic integrals of motion $K_i(p,q)$ because
\[
\{H_1,\omega_i\}=\{H_1,\Theta_i \bigl(K_i\bigr)\}=\Theta_i'\cdot\{H_1,K_i\}=0.
\]
So, the addition theorems (\ref{add-th}) could help us to classify algebraically superintegrable systems and vice versa. Plane curves with the genus $g\geq 1$ are related to
elliptic and Abelian integrals. Addition theorems of these functions are the content of Abel's theorem \cite{we08}.

The main result of this paper is that almost all the known examples of algebraically superintegrable systems relate with the one of the simplest addition theorem
\bq\label{add-ln}
e^x\e^y=\e^{x+y},\qquad\mathrm{or}\qquad \ln(x_1)+\ln(x_2)=\ln\left(x_1x_2\right)
\eq
associated with the zero-genus hyperelliptic curves $\mathcal C_j$
\bq\label{fr-1}
\mathcal C_j:\qquad \mu_j^2=P_j(\lambda_j)=f_j\lambda_j^2+g_j\lambda_j+h_j,\qquad j=1,\ldots,n,
\eq
where $f_j,g_j,h_j$ are linear functions on $n$ integrals of motion $H_1,\ldots,H_n$.

In fact,  if $S_{ij}(\lambda)=1 $  then after substitution (\ref{fr-1}) into (\ref{w-st}) one gets the sum of the rational functions
\[
\vartheta_j=\int\dfrac{1 }{\sqrt{g_j\lambda_j+h_j}}\,\mathrm d \lambda_j=\frac{\mu_j}{g_j}
\]
or logarithmic functions
\[
\vartheta_j=\int \dfrac{1}{\sqrt{f_j\lambda_j^2+g_j\lambda_j+h_j\,}}\,\mathrm d\lambda=
f_j^{-1/2}\,\ln\left(\mu_j+\dfrac{2f_j\lambda_j+g_j}{2\sqrt{f_j}}\right)\,.
\]
For the  hyperelliptic curves of higher genus one gets elliptic functions, which have more complicated addition law \cite{we08}.

In order to use addition low (\ref{add-ln}) we have to make the following steps:
\begin{itemize}
\item{We have to apply canonical transformation of the time that reduce $n$-th row of the St\"ackel matrix to the canonical Brill-Noether form \cite{ts99a,ts01}
\[S_{nj}=1,\qquad j=1,\ldots n,\]
such that
\bq\label{wk-n}
\omega_n=\dfrac{1}{N}\sum_{j=1}^n \int^{v_j(q_j)} \dfrac{1 }{\sqrt{f_j\lambda^2+g_j\lambda+h_j}}\,\mathrm d \lambda,
\eq
where $N$ is normalization, which is restored from $\{\omega_n,H_n\}=1.$
}
\item{Then we have to use addition law (\ref{add-ln}) for construction of the polynomial in momenta integrals of motion}
\item{We have to make inverse transformation of the time, which preserves the polynomial form of integrals such us it depends on $q$ variables only \cite{ts99a,ts01}}.
\end{itemize}

Let us consider  construction of  polynomial in the momenta integrals of motion  at $n=2$.

\subsubsection{Case  $f_1=f_2=0$}
If  $f_{1,2}=0$ we have
\bq\label{cal-ex}
\omega_2=\frac14\sum_{j=1}^2 \int^{v_j(q_j)} \dfrac{1 }{\sqrt{g_j\lambda+h_j}}\,\mathrm d \lambda= \frac{p_1{u_1}}{g_1}+\frac{p_2{u_2}}{g_2}=\frac{K}{g_1g_2},
\eq
where
\[K=g_2\,p_1{u_1}+g_1\,p_2{u_2}
\]
is the polynomial in the momenta integral of motion of the first or the third order. Remind that $g_{1,2}$ are linear functions on the St\"ackel integrals  $H_{1,2}$, which are the second order polynomials on $p_{1,2}$.

\begin{exam}

Let us consider the two-dimensional St\"{a}ckel system defined by two  Riemann surfaces
\[
\mathcal C_{1,2}:\qquad \mu^2=P_{1,2}(\lambda)=
(H_1\pm H_2)\lambda+\alpha_{1,2},
\]
and substitutions (\ref{fr-2})
\[
\mu_j=q_jp_j,\qquad\lambda_j=q_j^2\,,
\]
which give rise to the following separated equations
\[
p_{1,2}^2 = H_1\pm H_2+\frac{\alpha_{1,2}}{q_{1,2}^2}\,.
\]
Integrals of motion $H_1$ and $H_2$ are solutions of these separated equations
\bq\label{cal-int}
H_{1,2}=\dfrac{p_1^2\pm p_2^2}2-\dfrac{\alpha_1}{2q_1^2}\mp\dfrac{\alpha_2}{2q_2^2},
\eq
which coincide with integrals of motion for the two-particle Calogero system after an obvious point transformation
\bq\label{tr-c}
x=\frac{q_1-q_2}{2},\quad p_x=p_1-p_2,\qquad
y=\frac{q_1+q_2}{2},\quad p_y=p_1+p_2.
\eq
The angle variables (\ref{cal-ex}) read as
\ben
\omega_{1,2}&=& \dfrac{1}{4}\int^{q_1^2} \dfrac{1 }{\sqrt{(H_1+H_2)\lambda+\alpha_1}}\,\mathrm d \lambda
\pm  \dfrac{1}{4}\int^{q_2^2} \dfrac{1 }{\sqrt{(H_1-H_2)\lambda+\alpha_2}}\,\mathrm d \lambda\nn\\
\nn\\
&=&\dfrac{-p_1q_1(H_1-H_2)\mp p_2q_2(H_1+H_2)}{2(H_1+H_2)(H_1-H_2)}\,.\nn
\en
The corresponding cubic integral of motion $K$ (\ref{cal-ex}) is equal to
\[
K=2(H_1+H_2)(H_1-H_2)\omega_2=-p_1q_1(H_1-H_2)+p_2q_2(H_1+H_2).
\]
The bracket
\[
\{K,H_2\}=2(H_2^2-H_1^2)\,
\]
is easily restored from the canonical brackets (\ref{aa-br}).
\end{exam}

\subsubsection{Case  $f_1=k_1^2 f$ and  $f_2=k_2^2f$ with  integer $k_{1,2}$}
In this case, we have
\ben
\omega_2&=&\dfrac{1}{2}\sum_{j=1}^2 \int^{v_j(q_j)} \dfrac{1 }{\sqrt{f_j\lambda^2+g_j\lambda+h_j}}\,\mathrm d \lambda=
\sum_{j=1}^2 \dfrac{1}{2k_j\sqrt{f}}\ln\left( \dfrac{P_j'}{2k_j\sqrt{f}}+p_j{u_j}\right),\nn\\
&=&
\dfrac{1}{2k_1k_2\sqrt{f}}\,\ln \left[\left( \dfrac{P_1'}{2k_1\sqrt{f}}+p_1{u_1}\right)^{k_2}
\left( \dfrac{P_2'}{2k_2\sqrt{f}}+p_2{u_2}\right)^{k_1}\right],\nn
\en
where
\[P^{\,\prime}_{j}=\left.\frac{d P_{j}(\lambda)}{d\lambda}\right|_{\lambda=v_j(q_j)}=2f_jv_j(q_j)+g_j\,.\]
So function
\ben
\Phi_2&=& \exp(2k_1k_2\sqrt{f}\,\omega_2)=\left( \dfrac{P_1'}{2k_1\sqrt{f}}+p_1{u_1}\right)^{k_2}
\left( \dfrac{P_2'}{2k_2\sqrt{f}}+p_2{u_2}\right)^{k_1}=\nn\\
&=&\dfrac{1}{f^{(k_1+k_2)/2}}\left(K_{\ell}+\sqrt{f}\,K_m\right)\label{k5-gen}
\en
is the generating function of polynomial integrals of motion $K_m$ and $K_\ell$  of the $m$-th  and $\ell$-th order, respectively. The values of $m$ and $\ell$ depend on values of $k_{1,2}$.

As above, the algebra of  integrals $H_{1,2}$ and  $K_{m}$ may be restored from the canonical brackets (\ref{aa-br}).

\begin{exam}
Let us consider the two-dimensional St\"{a}ckel system defined by the two  Riemann surfaces
\[
\mathcal C_{1,2}:\qquad \mu_{1,2}^2=P_{1,2}(\lambda_{1,2})=k_{1,2}^2\lambda^2_{1,2}+\beta_{1,2}\lambda_{1,2}+(H_1\pm H_2)
\]
and substitutions (\ref{fr-2})
\[
\mu_j=\,p_j,\qquad\lambda_j=q_j\,,
\]
which give rise to the following separated equations
\[
p_{1,2}^2 = k_{1,2}^2 q_{1,2}^2+ \beta_{1,2}q_{1,2}+(H_1\pm H_2)\,.
\]
Integrals of motion $H_1$ and $H_2$ are solutions of these separated equations
\bq\label{os-int}
H_{1,2}=\dfrac{p_1^2 \pm p_2^2}2-\dfrac{k_1^2q_1^2\pm k_2^2q_2^2}{2}-\dfrac{\beta_1q_1\pm\beta_2q_2}{2},
\eq
which coincide with integrals of motion for the harmonic oscillator.
The same St\"ackel system coincides with the Kepler problem after well-known canonical transformation of the time, which changes the row of the St\"{a}ckel matrix  \cite{ts99a}.

The angle variable (\ref{cal-ex}) reads as
\ben
\omega_{2}&=&-\dfrac{1}{2}\left( \int^{q_j} \dfrac{1 }{\sqrt{k_1^2\lambda^2+\beta_1\lambda+(H_1+ H_2)}}\,\mathrm d \lambda -
\int^{q_2} \dfrac{1 }{\sqrt{k_2^2\lambda^2+\beta_2\lambda+(H_1-H_2)}}\,\mathrm d \lambda \right)=
\nn\\
&=&\dfrac{1}{2\sqrt{k_1^2k_2^2}}\left[\sqrt{k_2^2}\ln\left(p_1+\dfrac{P_1'}{2\sqrt{k_1^2}}\right)
-\sqrt{k_1^2}\ln\left(p_2+\dfrac{P_2'}{2\sqrt{k_2^2}}\right)
\right]\,.\nn
\en
Using suitable branches of $\sqrt{k_{1,}^2}\,$ one gets function (\ref{k5-gen})
\[
\Phi=\exp(2k_1k_2\omega_2)=\left(p_1-\dfrac{2k_1^2 q_1+\beta_1}{2k_1}\right)^{k_2}\left(p_2+\dfrac{2k_2^2q_2+\beta_2}{2k_2}\right)^{k_1},
\]
which generates polynomial integrals of motion $K_m$ and $K_\ell$. As an example, if  $k_1=1$ and
$k_2=3$ one gets third and fourth order integrals of motion  $K_m$ and $K_\ell$ respectively.
\end{exam}

Of course, oscillator is one of the well-studied superintegrable systems, which was earmarked to illustrate  generic construction only. Additional theorem (\ref{add-ln}) allows us to get a huge family  of the $n$-dimensional superintegrable systems, which have to be classified and studied.

\subsection{Classification}
In order to classify superintegrable systems  associated with the addition theorem (\ref{add-ln}) we have to start with a pair of the Riemann surfaces
\bq\label{dr-eq}
\mathcal C_{j}:\qquad\mu^2=P_j(\lambda)= f\lambda^2+g_j\lambda+h_j,\qquad j=1,2,
\eq
where
\[f=\alpha H_1+\beta H_2+\gamma,\quad
  g_j=\alpha^g_jH_1+\beta^g_j H_2+\gamma^g_j,\quad
  h_j=\alpha^h_jH_1+\beta^h_j H_2+\gamma^h_j,\]
and $\alpha$, $\beta$ and $\gamma$ are real or complex numbers.

In order to use addition theorem (\ref{add-ln}) we  fix last row of the St\"ackel matrix. Namely, substituting
\bq\label{m-rest}
S_{2j}(\lambda)=\kappa_j
\eq
into the (\ref{stc}-\ref{w-st}) one gets
\bq\label{th-eq}
\vartheta_{2j}=\int \dfrac{S_{2j}(q_j)\,\mathrm dq_j}{p_j}=
\int\dfrac{\kappa_j\,\mathrm d\lambda_j}{\sqrt{P_j}}=
\kappa_jf^{-1/2}\,\ln\left(\mu_j+\dfrac{2f\lambda_j+g_j}{2\sqrt{f}}\right)
\,,
\eq
so the angle variable
\[
\omega_2=\dfrac{1}{\sqrt{f}}\,\ln \left[\left(p_1{u_1}+ \dfrac{P_1'}{2\sqrt{f}}\right)^{\kappa_1}
\left(p_2{u_2}+\dfrac{P_2'}{2\sqrt{f}}\right)^{\kappa_2}\right],
\qquad
P^{\,\prime}_{j}=\left.\frac{d P_{j}(\lambda)}{d\lambda}\right|_{\lambda=v_j(q_j)},\]
is the multi-valued function on the desired algebraic argument
\[
K=\left(p_1{u_1}+ \dfrac{P_1'}{2\sqrt{f}}\right)^{\kappa_1}
\left(p_2{u_2}+\dfrac{P_2'}{2\sqrt{f}}\right)^{\kappa_2}.
\]
If $\kappa_{1,2}$ are positive integer, then
\bq\label{k-kappa}
K=\left(\dfrac{1}{2\sqrt{f}}\right)^{\kappa_1+\kappa_2}\left(K_{\ell}+\sqrt{f}\,K_m\right)
\eq
is the generating function of polynomial integrals of motion $K_m$ and $K_\ell$  of the $m$-th  and $m\pm 1$-th order in the momenta, respectively.

As an example we have
\bq\label{k-m}
\begin{array}{ll}
K_m=2\,(p_1{u_1}\,P'_2+p_2{u_2}\,P'_1),\quad &\kappa_1=1,\, \kappa_2=1,\\
K_m=2P'_2\,\Bigl(2p_2u_2\,P'_1+p_1u_1\,P'_2\Bigr)+8p_1u_1p_2^2u_2^2f\,,
\quad&\kappa_1=1,\,\kappa_2=2,\\
K_m=2{P'_2}^2\Bigl(3p_2u_2P'_1+p_1u_1P'_2\Bigr)+8p_2^2u_2^2\Bigl(p_2u_2P'_1+3p_1u_1P'_2\Bigr)f,
\quad&\kappa_1=1,\,\kappa_2=3,
\end{array}
\eq
where $m=1,3$, $m=3,5$ ¨ $m=3,7$, because $P'_{1,2}$ and $f$ are linear functions on  $H_{1,2}$, which are the second order polynomials on momenta.

The corresponding expressions for the $K_\ell$ look like
\bq\label{k-ell}
\begin{array}{ll}
K_{\ell}=P'_1P'_2+4p_1p_2{u_1u_2}f,\quad &\kappa_1=1,\, \kappa_2=1,\\
K_{\ell}=P'_1{P'_2}^2+4fp_2u_2(p_2u_2P'_1+2p_1u_1P'_2)\,,
\quad&\kappa_1=1,\,\kappa_2=2,\\
K_{\ell}=P'_1{P'_2}^3+12P'_2p_2u_2(P'_2p_1u_1+P'_1p_2u_2)f+16p_1u_1p_2^3u_2^3f^2\,.
\quad&\kappa_1=1,\,\kappa_2=3.
\end{array}
\eq

The imposed condition (\ref{m-rest}) leads to some restrictions on the functions $v_j(q_j)$ and $u_j(q_j)$. In fact, substituting canonical variables (\ref{fr-2}) into the equations (\ref{dr-eq})
we obtain  the following expression for the St\"{a}ckel matrix
\bq\label{s-fin}
S=\left(\begin{array}{cc}
\dfrac{\alpha v_1^2+\alpha^g_1 v_1+\alpha^h_1}{u_1^2} & \dfrac{\alpha v_2^2+\alpha^g_2 v_2+\alpha^h_2}{u_2^2}\\
\\
\dfrac{\beta v_1^2+\beta^g_1 v_1+\beta^h_1}{u_1^2}& \dfrac{\beta v_2^2+\beta^g_2 v_2+\beta^h_2}{u_2^2}
\end{array}\right)\,,\qquad \det S\neq 0.
\eq
So,  for a given $\kappa_{1,2}$ expressions for $\vartheta_{2j}$ (\ref{th-eq}) yield two differential equations on functions $u,v$ and parameters $\beta$:
\bq\label{uv-eq}
S_{2j}(q_j)=\dfrac{\kappa_j\,v'_j(q_j)}{u_j(q_j)}\quad \Longrightarrow\quad \kappa_j u_jv'_j=\beta v_j+\beta^g_jv_j+\beta^h_j,\qquad j=1,2.
\eq
For the St\"ackel systems with rational or trigonometric metrics, we have to solve these equations in the space of the truncated Laurent or Fourier polynomials, respectively.

\begin{prop}
If $\kappa_{j}\neq 0$ equations (\ref{uv-eq}) have the following three monomial solutions
\bq \label{uv-sol}
\begin{array}{llll}
\mathrm I\qquad&\beta=0,\quad \beta_j^h=0,\qquad& u_j=q_j,\qquad &v_j=q_j^{\frac{\beta_j^g}{\kappa_j}},
\\
\\
\mathrm{II}\qquad&\beta_j^g=0,\quad\beta_j^h=0,\qquad &u_j=1,\qquad& v_j=-\kappa_j(\beta\, q_j)^{-1},\\
\\
\mathrm{III}\qquad&\beta=0,\quad\beta_j^g=0,\qquad &u_j=1,\qquad &v_j=\kappa_j^{-1}\beta_j^h\, q_j,
\end{array}
\eq
up to canonical transformations. The fourth solution $\mathrm{(\,IV)}$ is the combination of the  first and third solutions for the different $j$'s.
\end{prop}
In order to prove this fact we  can substitute $u=aq^m$ and $v=bq^k$ into the (\ref{uv-eq}) and divide resulting equation on $q^{m+k-1}$
\[
abk\kappa = b^2\beta q^{k+1-m}+q^{1-m}b\beta^g+q^{1-k-m}\beta^h.
\]
Finally, we differentiate it by $q$  and multiply on $q^{m}$
\[
0=-(m-k-1)b^2\beta q^{k}-(m-1)b\beta^g -(m+k-1)\beta^h\,q^{-k}.
\]
Such as $b\neq 0$ and $k\neq 0$ one gets three solutions (\ref{uv-sol}) only.

Then we suppose that after some point transformation
\bq \label{z-dr}
\begin{array}{ll}
x=z_1(q), &  y=z_2(q),\\
p_x=\mathrm w_{11}(q)p_1+\mathrm w_{12}(q)p_2,\qquad &p_y=\mathrm w_{21}(q)p_1+\mathrm w_{22}(q)p_2,\end{array}
\eq
where $\mathrm w_{ij}\neq 0$, kinetic part of the Hamilton function $H_1=T+V$ has a special form
\[
T=\sum\left(S^{-1}\right)_{1j}p_j^2=\mathrm{g}_{11}(x,y)p_x^2
+\mathrm{g}_{12}(x,y)p_xp_y+\mathrm{g}_{22}(x,y)p_y^2,
\]
where $\mathrm{g}$ is a metric on a configurational manifold. For instance, if we suppose that
\[T=\sum\left(S^{-1}\right)_{1j}p_j^2=p_xp_y,
\]
then  one gets the following  algebraic equations
\bq\label{alg-dr}
\mathrm w_{11}\mathrm w_{21}=\left(S^{-1}\right)_{11},\qquad
\mathrm w_{12}\mathrm w_{21}+ \mathrm w_{11}\mathrm w_{22}=0,\qquad
\mathrm w_{12}\mathrm w_{22}=\left(S^{-1}\right)_{12}
\eq
 and  the  partial differential equations
\bq\label{pde-dr}
\{x,p_x\}=\{y,p_y\}=1,\qquad \{p_x,y\}=\{p_y,x\}=\{p_x,p_y\}=0.
\eq
on  parameters $\alpha$ and  functions  $z_{1,2}(q_1,q_2)$, $\mathrm w_{kj}(q_1,q_2)$.

The remaining free parameters $\gamma,\gamma_j^h,\gamma_j^g$ determine the corresponding potential part of the Hamiltonian $V(x,y)$. In fact, since integrals $H_{1,2}$ is defined up to the trivial shifts $H_k\to H_k+c_k$, our potential $V(x,y)$ depends on three arbitrary parameters only.

Summing up, in order to get all the superintegrable systems on a complex Euclidean space $E_{2}(\mathbb{C})$  associated with the addition theorem (\ref{add-ln}) we have to solve equations (\ref{uv-eq},\ref{alg-dr},\ref{pde-dr}) with respect to functions
$u_{j}(q_j)$, $v_{j}(q_j)$, $z_{1,2}(q_1,q_2)$, $\mathrm w_{kj}(q_1,q_2)$ and parameters $\alpha$ and $\beta$.

\begin{exam}
Let us consider second solution from the list (\ref{uv-sol}) at $\kappa_1=1$ and $\kappa_2=2$.  In this case equations (\ref{alg-dr},\ref{pde-dr}) have the following partial solution
\[
S=\left(\begin{array}{cc}1\quad& 0\\ \frac{1}{q_1^2}\quad& \frac{4}{q_2^2}\end{array}\right),\qquad
\begin{array}{ll}
x={\sqrt{q_1}q_2},\quad &
p_x= \frac{\sqrt{q_1}}{q_2}\,p_1+\frac{1}{2\sqrt{q_1}}\,p_2\,,\\
y=\frac{\sqrt{q_1}}{q_2},\quad&
p_y= \sqrt{q_1}q_2\,p_1-\frac{q_2^2}{2\sqrt{q_1}}\,p_2\,.\end{array}
\]
Adding potential terms one gets two  Riemann surfaces
\[
\mathcal C_{1}:\quad
\mu^2={H_2}{\lambda^2}-{H_1}{\lambda}-{\gamma_1},\quad\mathrm{and}\quad
\mathcal C_{2}:\quad\mu^2={H_2}{\lambda^2}-{2\gamma_2}{\lambda}+{4\gamma_3}
\]
where $\mu_j=\,p_j$, $\lambda_j=q_j^{-1}$. Solutions of the corresponding separated variables
are the  second order St\"akel integrals of motion, which in physical variables  look like
\bq\label{h1-5}
H_1=p_xp_y+\gamma_1 xy +\dfrac{\gamma_2}{\sqrt{xy^3}}+\dfrac{\gamma_3}{y^2},\qquad
H_2=\dfrac{(p_xx-p_yy)^2}4-\gamma_2\sqrt{\dfrac{x}{y}}-\gamma_3\dfrac{x}{y}\,.
\eq
It is new integrable systems, which is missed in the known lists of superintegrable systems \cite{cd06,mw07,ran97, ran01}.

In this case  integrals of motion $K_5$ (\ref{k-m}) and $K_6$ (\ref{k-ell})
are the fifth and sixth order polynomials in the momenta, respectively.
Of course, we can try to get quartic, cubic and quadratic integrals of motion $K_{4}$, $K_3$ and $K_{2}$ from the recurrence relations
\bq\label{rec-H}
K_5=\{K_4,H_2\},\qquad
K_4=\{K_3,H_2\},\qquad
K_3=\{K_2,H_2\}
\eq
and the equation $\{H_1,K_{j}\}=0$, $j=4,3,2$.

Solving first recurrence equation one gets quartic integral of motion
\[
K_4=-4(xp_x-yp_y)^2(p_x^2+\gamma_1 y^2)-\frac{4\gamma_2^2}{xy}
+\frac{8p_x(xp_x-yp_y)\gamma_2}{\sqrt{xy}}
+\frac{16x(p_x^2+\gamma_1y^2)\gamma_3}{y}.
\]
The other recurrence relations (\ref{rec-H}) have not polynomial solutions.
Of course, we can try modify recurrence relations
\[
\{K_3,H_2\}=K_4+F_4(H_1,H_2),\qquad \{K_2,H_2\}=K_3+F_3(H_1,H_2)\,
\]
in order to get cubic and quadratic integrals of motion. However, using ansatz
\[
K_2=h_1(x,y)p_x^2+h_2(x,y)p_xp_y+h_3(x,y)p_y^2+h_4(x,y)
\]
we can directly prove that there is not the additional quadratic integral of motion, which commute with $H_1$ (\ref{h1-5}).
\end{exam}

\section{The Drach systems}
\setcounter{equation}{0}
In 1935 Jules Drach published two articles on the Hamiltonian systems with the third order integrals of motion on a complex Euclidean space $E_{2}(\mathbb{C})$  with the following Hamilton function \cite{dr35}
\bq\label{ham-dr}
H_1=p_xp_y+U(x,y)\,.
\eq
Up to canonical transformations $x\to a x$ and $y\to b y$ the corresponding potentials look like
\cite{ran97,ts00}:
\ben
(a)\qquad U&=&\dfrac{\alpha}{xy}+\beta x^{r_1}y^{r_2}+\gamma
x^{r_2}y^{r_1}\,,\quad\mbox{\rm where}\quad r_j^2+3r_j+3=0\,,
\nn\\
(b)\qquad U&=&\dfrac{\alpha }{\sqrt{xy}}+\dfrac{\beta}{(y-x)^2}+\dfrac{\gamma\,(y+x)}{\sqrt{xy}\,(y-x)^2}\,,
\nn\\
(c)\qquad
U&=&\alpha \,xy+\dfrac{\beta}{(y-x)^2}+\dfrac{\gamma} {(y+x)^2}\,,
\nn\\
(d)\qquad U&=&\dfrac{\alpha }{\sqrt{y(x-1)\,}}+\dfrac{\beta}{\sqrt{y(x+1)\,}}
+\dfrac{\gamma x}{\sqrt{x^2-1^2\,}}\,,
\nn\\
(e)\qquad U&=&\dfrac{\alpha }{\sqrt{xy\,}}+\dfrac{\beta}{\sqrt{x\,}}
+\dfrac{\gamma} {\sqrt{y\,}}\,,\nn\\
(f)\qquad U&=&\alpha \,xy +\beta y\dfrac{2x^2+1}{\sqrt{x^2+1\,}}+\dfrac{\gamma x}{\sqrt{x^2+1\,}}\,,\nn\\
(g)\qquad
U&=&\dfrac{\alpha }{(y+x)^2}+\beta(y-x)+\dfrac{\gamma(3y-x)(y-3x)}3\,,
\nn\\
(h)\qquad U&=&(y+\dfrac{mx}3)^{-2/3}\,\left[\alpha+\beta\,(y-mx/3)+
\gamma\,(y^2-\dfrac{14mxy}3+\dfrac{m^2x^2}9)\right]\,,
\nn\\
(k)\qquad U&=&\alpha  y^{-1/2} +\beta x y^{-1/2} +\gamma x\,,
\nn\\
(l)\qquad U&=&\alpha \left(y-\dfrac{\rho x}3\right)+\beta x^{-1/2} +\gamma x^{-1/2}(y-\rho x)\,.
\nn
\en
Such as Drach made some assumptions on the form of the third order integrals of motion $K_3$ (\ref{k3-dr}) in the calculation it is not immediately clear whether the obtained list is complete.

\subsection{Non-separable systems.}
Let us discuss the Drach systems, which can not be reduced to the St\"ackel systems by any point  transformation of variables.

The first system (a) is non-St\"ackel system related to the three-particle periodic Toda lattice in the center-of-mass frame \cite{ts00} and there are global action-angle variables  \cite{hk08}.

The (h) system is reduced to the St\"ackel system by non-point canonical transformation and, therefore, existence of the third order integral of motion is related with this non-point transformation \cite{ts00}. Later this system has been rediscovered by Holt \cite{ho82}.

For the (k) case in the  Drach papers \cite{dr35} we can find Hamiltonian
\bq\label{hkd}
H_1^{(k)}=p_xp_y+\alpha  y^{-3/2} +\beta x y^{-3/2} +\gamma x
\eq
and the following cubic integral of motion
\bq\label{kkd}
K_3^{(k)}=6w(x,y)\,\left(\dfrac{\partial H}{\partial x}\,p_y
-p_x\,\dfrac{\partial H}{\partial y}\right) -P(p_x,p_y,x,y),\eq
where
\[
P=3p_x^2p_y\,,\qquad w=-y.
\]
It is easy to prove that $\{H_1^{(k)},K_3^{(k)}\}\neq 0$ and, therefore, we have to 
suggest the possibility of a small mistake in the Drach papers \cite{dr35}.

Following to \cite{ts00}, if we  solve equation $\{p_xp_y+U(x,y),K_3^{(k)}\}=0$
with respect to $U(x,y)$,
 then one gets our case (k) 
\[
H_1=p_xp_y+\alpha  y^{-1/2} +\beta x y^{-1/2} +\gamma x.
\]
On the other hand, we have proven directly that the
Hamiltonian $H_1^{(k)}$ (\ref{hkd}) has  one second order integral of motion
\bq\label{h2k}
H_2^{(k)}=p_x^2-4\beta y^{1/2}+2\gamma y
\eq
and has not cubic integral of motion. Moreover, it is easy to see that quadratic integrals of motion $H_{1,2}^{(k)}$ (\ref{hkd},\ref{h2k}) can not be reduced to the St\"ackel integrals by any point  transformation of variables.

 Below we do not consider (a) and (h) systems and consider (k) case in our notation only.

\subsection{Classification.}

For the Drach systems $\kappa_1=\kappa_2=\pm1,1/2$ and
\ben
\Phi_2&=& \exp(2\sqrt{f}\,\omega_2)=\left( \dfrac{P_1'}{2\sqrt{f}}+p_1{u_1}\right)\left( \dfrac{P_2'}{2\sqrt{f}}+p_2u_2\right)=\nn\\
&=&
\dfrac{1}{4f}\left(K_{\ell}+\sqrt{f}\,K_m\right)\label{phi-ln}
\en
may be considered as the generating function of the polynomial integrals of motion (\ref{k-m}-\ref{k-ell})
\ben
K_m&=&2\Bigl(p_1{u_1}\,P'_2+p_2{u_2}\, P'_1\Bigr),\quad m=1,3,
\label{k3-dr}\\
K_{\ell}&=&P'_1P'_2+4p_1p_2{u_1u_2}f,\qquad \ell=2,4,\nn
\en
 of the $m$-th  and $\ell$-th order, respectively. It's clear that $m=1,3$ and $\ell=2,4$, because $P'_{1,2}$ and $f$ are linear functions on  $H_{1,2}$, which are  second order polynomials on momenta.

 We have to underline, that we use different $\kappa_{1}=\kappa_2{2}=\pm1,1/2$ for the agreement of $K_m$ (\ref{k3-dr}) with the initial Drach integrals of motion \cite{dr35} only.

 One gets   third order polynomial integral of motion $K_3$ (\ref{k3-dr})  if and only if $P'_1(\lambda)$ or $P'_2(\lambda)$ depends on $H_1$ or $H_2$. It leads to the additional restrictions on $\alpha$'s and $\beta$'s
\bq\label{cub-dr}
\sum_{k=1}^2\dfrac{\partial^2 P_j(\lambda)}{\partial H_k\partial \lambda}\neq 0 \qquad\textrm{for}\qquad j=1
\quad\textrm{or}\quad j=2.
\eq

In order to get all the superintegrable systems on a complex Euclidean space $E_{2}(\mathbb{C})$  associated with addition theorem (\ref{add-ln}) we have to solve equations (\ref{uv-eq},\ref{alg-dr},\ref{pde-dr}) and (\ref{cub-dr})
at $\kappa_{1,2}=1$ with respect to the functions
$u_{j}(q_j)$, $v_{j}(q_j)$, $z_{1,2}(q_1,q_2)$, $\mathrm w_{kj}(q_1,q_2)$ and parameters $\alpha$ and $\beta$.

\begin{prop}
The Drach list of the St\"ackel systems with the cubic integral of motion   (\ref{k3-dr}) associated with the addition theorem (\ref{add-ln}) is complete up to canonical transformations of the extended phase space.
\end{prop}
The results of corresponding calculations may be joined into the table:
\vskip0.5truecm
\par\noindent
\begin{tabular}{|c|c|c|c|c|}
\hline
&  $\mathcal C_{1,2}$ (\ref{dr-eq}) & subs. (\ref{fr-2})& $z_{1,2}$ (\ref{z-dr})& S   \\
\hline
    b & $\begin{array}{c}\\
        \mu^2=H_1\lambda^2+(H_2+2\alpha)\lambda-\beta+2\gamma\\
       \\
        \mu^2=H_1\lambda^2+(H_2-2\alpha)\lambda-\beta-2\gamma\\
       \end{array}$
       & $\begin{array}{c} \\  \\ \mu_j=p_jq_j\end{array}$ & $z_{1,2}=\frac{(q_1\pm q_2)^2}4$  &$\left(\begin{smallmatrix} q_1^2&q_2^2\\1&1\end{smallmatrix}\right)$\\
\cline{1-2} \cline{4-5}
    c & $\begin{array}{c}\\
        \mu^2=\dfrac{\alpha}4\lambda^2+\left(H_2+\dfrac{H_1}2\right)\lambda+\gamma\\
       \\
        \mu^2=\dfrac{\alpha}4\lambda^2+\left(H_2-\dfrac{H_1}2\right)\lambda-\beta\\
       \end{array}$
       & $  \begin{array}{c}\lambda_j=q_j^2\\ \\ \kappa_j=\frac{1}{2}\end{array}$ & $z_{1,2}=\frac{q_1\pm q_2}{2}$ &$\left(\begin{smallmatrix} \frac12&-\frac12\\1&1\end{smallmatrix}\right)$ \\
\hline
    d & $\begin{array}{c}\\
        \mu^2=H_2\lambda^2-\sqrt{8}(\alpha+\beta)\lambda+H_1-2\gamma\\
       \\
        \mu^2=H_2\lambda^2-\sqrt{8}(\alpha-\beta)\lambda+H_1+2\gamma\\
       \end{array}$
       & $\begin{array}{c} \\ \\  \mu_j=p_j \end{array}$ & $\begin{array}{l}z_1=\frac{q_1^2+q_2^2}{2q_1q_2}\\ \\ z_2= q_1q_2\end{array}$ &$\left(\begin{smallmatrix} 1&1\\ \frac1{q_1^2}&\frac{1}{q_2^2}\end{smallmatrix}\right)$ \\
\cline{1-2} \cline{4-5}
f & $\begin{array}{c}\\
        \mu^2=H_2\lambda^2-\left(\dfrac{\gamma}2-H_1\right)\lambda-\dfrac{\alpha}4-\dfrac{\beta}2\\
       \\
        \mu^2=H_2\lambda^2-\left(\dfrac{\gamma}2+H_1\right)\lambda-\dfrac{\alpha}4+\dfrac{\beta}2\\
       \end{array}$
       & $\begin{array}{c}  \lambda_j=q_j^{-1}\\ \\ \kappa_j=-1\end{array}$ &$\begin{array}{l}z_1=\frac{q_1-q_2^2}{2\sqrt{q_1q_2}}\\  \\ z_2= \sqrt{q_1q_2}\end{array}$  &$\left(\begin{smallmatrix} \frac1{q_1}&\frac{-1}{q_2}\\ \frac1{q_1^2}&\frac{1}{q_2^2}\end{smallmatrix}\right)$\\
\hline
e & $\begin{array}{c}\\
        \mu^2=H_1\lambda^2+2(\beta+\gamma)\lambda+H_2+2\alpha\\
       \\
        \mu^2=H_1\lambda^2-2(\beta-\gamma)\lambda+H_2-2\alpha\\
       \end{array}$
       & $\begin{array}{c} \\ \\  \mu_j=p_j\end{array}$ & $z_{1,2}=\frac{(q_1\pm q_2)^2}4$ &$\left(\begin{smallmatrix} q_1^2&q_2^2\\1&1\end{smallmatrix}\right)$ \\
\cline{1-2} \cline{4-5}
k & $\begin{array}{c}\\
        \mu^2=\dfrac{\gamma}{2}\lambda^2+(\beta+H_1)\lambda+H_2+\alpha\\
       \\
        \mu^2=\dfrac{\gamma}{2}\lambda^2+(\beta-H_1)\lambda+H_2-\alpha\\
       \end{array}$
       & $\begin{array}{c} \lambda_j=q_j \\ \\ \kappa_j=1\end{array}$ & $\begin{array}{l} z_1=\frac{q_1-q_2}{2}\\ \\ z_2=\frac{(q_1+q_2)^2}{4}\end{array}$
       &$\left(\begin{smallmatrix} q_1&-q_2\\1&1\end{smallmatrix}\right)$
       \\
\hline
g & $\begin{array}{c}\\
        \mu^2=-\dfrac{\gamma}{3}\lambda^2+\left(\dfrac{H_1}2+H_2\right)\lambda+\alpha\\
       \\
        \mu^2=-\dfrac{\gamma}{3}\lambda^2-\dfrac{\beta}4\lambda+\dfrac{H_2}{4}-\dfrac{H_1}8\\
       \\
       \end{array}$
       & $\begin{array}{c}
           \scriptstyle \mu_1=p_1q_1,\,\,\lambda_1=q_1^2\\
           \\
           \scriptstyle \mu_2=2p_1,\,\,\lambda_2=q_2\\
           \\
           \scriptstyle \kappa_j=\frac{1}{2}
          \end{array}$ & $z_{1,2}=\frac{q_1\pm q_2}{2}$
           &$\left(\begin{smallmatrix} \frac12&-\frac12\\1&1\end{smallmatrix}\right)$\\
\hline
\end{tabular}
\vskip0.5truecm
\par\noindent
Similar to the oscillator and the Kepler problem,  the Kepler change of the time $t\to \widetilde{t}$, where
\bq\label{t-change}
d\widetilde{t}=v(q)dt,\qquad v(q)=\dfrac{\det{S}}{\det\widetilde{S}},
\eq
relates the Drach  systems  (b),(d) and (e) with the systems (c),(f) and (k), respectively.
Here $S$ are the St\"ackel matrices for (b),(d) and (e) systems and $\widetilde{S}$ are the St\"ackel matrices for (c),(f) and (k) systems. Remind, that these matrices $S$ and $\widetilde{S}$ have different first row only,  see \cite{ts99a,ts01}.

\subsection{Integrals of motion}
Using definitions (\ref{k3-dr}) we can prove that integral of motion $K_4$ (\ref{k3-dr}) is the function on  $H_1,H_2$ and  $K_3$
\bq
K_4^2=16h_1h_2f^2+(K_3^2-4h_1g_2^2-4h_2g_1^2)f+g_1^2g_2^2.
\eq
Substituting this expression into the definition of $\Phi_2$ (\ref{phi-ln}) we can get integral $K_3$ as function on  the action-angle variables $I_{1,2}=H_{1,2}$ and $\omega_2$.

As usual, polynomial algebra of integrals of motion $H_{1,2}$ and $K_3$ follows from the canonical brackets (\ref{aa-br})
\ben
\{H_1,H_2\}&=&\{H_1,K_3\}=\{H_1,K_4\}=0,\nn\\
\{H_2,K_3\}&=&\delta K_4,\qquad \{H_2,K_4\}=\delta fK_3,\label{alg-b}\\
\{K_3,K_4\}&=&F_Z(H_1,H_2,K_3)\,,\nn
\en
where
 \begin{itemize}
       \item $\delta=\phantom{-}4$, ~$F_{I}=16f(g_1h_2+g_2h_1)-4g_1g_2(g_1+g_2)$\phantom{abs,} for b,c cases;
       \item $\delta=-2$, ~$F_{II}=K_3+32fh_1h_2-4(g_1^2h_2+g_2^2h_1)$\phantom{absde} for d,f cases;
       \item $\delta=\phantom{-}2$, ~$F_{III}=4f(g_1^2+g_2^2)-16f^2(h_1+h_2)$\phantom{absdefx,} for e,k cases;
       \item $\delta=\phantom{-}4$, ~$F_{IV}=2f(g_1^2+8g_1h_2)-8f^2h_1-4g_1g_2^2$\phantom{absd,} for g case.
\end{itemize}
The difference in the values of $\delta$ is related with the difference in $\kappa$'s which has been defined by the Drach integrals of motion \cite{dr35}. Of course, we can put $\kappa_1=\kappa_2=1$ in all the cases such that $\delta=2$ and  polynomials $F$ look as in (\ref{alg-comm}). It reduces polynomials $P_{1,2}$ in the table only.

As above (\ref{rec-H}), we can try to find another second order polynomial integral of motion $K_2$ from the equations
\[
K_3=\{H_2,K_2\},\qquad \{H_1,K_2\}=0.
\]
Solutions of these equations
\bq
K_2=\pm\Bigl(2p_1p_2u_1u_2+2v_1v_2f+v_1g_2+v_2g_1\Bigr)=\pm\dfrac{K_4-g_1g_2}{2f}\label{int-k2}
\eq
have been found in \cite{ran97} in framework of the Lagrangian formalism. The algebras of quadratic integrals of motion $H_{1,2}$ and $K_2$ have been considered in \cite{cd06}.

There is some opinion that all superintegrable systems with quadratic (linear) integrals of motion are multiseparable, i.e allows the separation of variables in the Hamilton-Jacobi equation in at least two different coordinate systems on the configuration space \cite{mw07}. The some of the Drach systems may be considered as  counterexamples associated with the Lie surfaces \cite{cd06}.

Namely, three of the superintegrable Drach systems with quadratic integrals $H_1,H_2,K_2$ are separable in the one coordinate system on the configuration space only.
\begin{prop}
For the (b) and (c) Drach systems integrals of motion $H_1,K_2$ are separable in the coordinates
\[
x=\frac{q_2}{2q_1},\qquad y=q_1q_2,\]
and the corresponding separated relations do not allows us to get cubic integrals of motion.

For the (e) and (g) cases integrals of motion $H_1,K_2$ are separable in the coordinates
\[
x=\frac{q_1-q_2}{2},\quad y=-\frac{(q_1+q_2)^2}{4}
\]
and we can use the corresponding separated relations to the construction of the cubic integrals of motion.

For the (d), (f) and (k) cases   quadratic integrals of motion $H_1$ and $K_2$ do not separable by the point transformations.
\end{prop}
\par\noindent
We can prove this Proposition by using computer program from \cite{ts05}.

\subsection{The (l) system.}
Without lost of generality we can put $\rho=-3$ in the (l) case. Substituting this Hamiltonian
into the computer program from \cite{ts05} one gets the separated variables
\[
x=\dfrac{(q_1-q_2)^2}2,\qquad y= \dfrac{(q_1+q_2)^2}{2}
\]
and the corresponding separated relations
\bq\label{sep-rel-l}
p_j^2=P_j(q_j)= -4\alpha q_j^4\mp 8\sqrt{2}\gamma q_j^3+4H_1q_1^2\mp 4\sqrt{2}\beta q_j+H_2,\qquad j=1,2.
\eq
which give rise to one hyperelliptic curve $\mu^2=P(\lambda)$  at $\mu=p_j$ and $\lambda=\pm q_j$.

The angle variable
\ben
\omega_2&=&\dfrac12\int^{q_1} \dfrac{\mathrm d \lambda}{\sqrt{P(\lambda)}}+\dfrac12
\int^{q_2} \dfrac{\mathrm d \lambda}{\sqrt{P(-\lambda)}}\nn\\
&=&\dfrac12\int^{q_1} \dfrac{\mathrm d \lambda}{\sqrt{P(\lambda)}}-\dfrac12
\int^{-q_2} \dfrac{\mathrm d \lambda}{\sqrt{P(\lambda)}}
\en
is a sum of the incomplete elliptic integrals of the first kind on the common hyperelliptic curve. According to \cite{we08,dr35} there is addition theorem and additional cubic integral of motion
\[
 {K}_3=2(\widetilde{P}'_1\,p_2+\widetilde{P}'_2\,p_1),
\]
which looks like as the Drach integral (\ref{k3-dr}),  but in this case functions
\[\widetilde{P}'_{1,2}=(q_1+q_2)^2\dfrac{\partial}{\partial q_{1,2}}\, \dfrac{P(\pm q_{1,2})}{(q_1+q_2)^4}\,\]
have completely another algebro-geometric explanation. All the details will be published in the forthcoming publications.

As sequence, the algebra of integrals of motion $H_{1,2}$ and $K_3$ differs from the corresponding algebras for other Drach systems related with another addition theorem. As an example, the recurrence chain $ {K}_{j+1}=\{H_2, {K}_j\}$ terminates on the fourths step only
\[
 {K}_7=\{H_2, {K}_6\}=-480 {K}_4 {K}_3 +256H_1^2 {K}_3 -768\alpha H_2 {K}_3-3072\beta\gamma {K}_3\,.
\]

The solution of the inverse recurrence chain looks like
\[
 {K}_2= p_y^2+2\alpha x-4 \gamma \sqrt{x}.
\]
It is interesting, that  the algebra of quadratic integrals of motion  $H_{1,2}$ and $K_2$ is one of the standard cubic algebras  \cite{cd06} and we do not explanation of this fact.

\section{New superintegrable systems on zero-genus hyperelliptic curves at $\kappa_1=1$ and $\kappa_2=2,3$}

\setcounter{equation}{0}

Let us put $\kappa_1=1$ and $\kappa_2=2$ in (\ref{uv-eq}) and try to solve  equations (\ref{alg-dr})-(\ref{pde-dr}). Here is one superintegrable system with cubic additional integral $K_m$ (\ref{k-m}) and quadratic integral $K_\ell$ (\ref{k-ell})
\[
V_{III}= \gamma_1(3x+y)(x+3y)+{\gamma_2}{(x+y)}+{\gamma_3}{(x-y)}\,,\qquad
S=\left(
  \begin{smallmatrix}
    a & b \\
    1 & 1
  \end{smallmatrix}
\right),
\]
and seven systems with the real potentials
\[
\begin{array}{ll}
V_{I}= \gamma_1(3x+y)(x+3y)+\dfrac{\gamma_2}{(x+y)^2}+\dfrac{\gamma_3}{(x-y)^2},
\qquad
&S=\left(
  \begin{smallmatrix}
    a/q_1 & b/q_2 \\
    1/q_1 & 2/q_2
  \end{smallmatrix}
\right),\nn\\
V_{II}^{(1)}= \gamma_1 xy +\dfrac{\gamma_2}{\sqrt{x^3y}}+\dfrac{\gamma_3}{x^2}\,,
\qquad
&S=\left(
  \begin{smallmatrix}
    1/q_1 & 1 \\
    1/q_1^2 & 4/q_2^2
  \end{smallmatrix}
\right),\nn\\
V_{II}^{(2)}= \dfrac{\gamma_1}{\sqrt{xy}}+\dfrac{\gamma_2}{x^2} +\dfrac{\gamma_3 y}{x^3}\,,
\qquad
&S=\left(
  \begin{smallmatrix}
    0 & 1 \\
    1/q_1^2 & 4/q_2^2
  \end{smallmatrix}\right),\nn\\
V_{II}^{(3)}= \dfrac{\gamma_1}{\sqrt{xy}}+\dfrac{\gamma_2}{\sqrt{x^3y}}
+\dfrac{\gamma_3}{x^{5/4}y^{3/4}}\,,
\qquad
&S=\left(
  \begin{smallmatrix}
    1 & 0 \\
    1/q_1^2 & 4/q_2^2
  \end{smallmatrix}\right),\nn\\
V_{II}^{(4)}= \gamma_1 xy+\dfrac{\gamma_2 y}{x^3}+\dfrac{\gamma_3 y^3}{x^5}\,,
\qquad
&S=\left(
  \begin{smallmatrix}
      0 & 4/q_2 \\
    1/q_1^2 & 4/q_2^2
  \end{smallmatrix}\right),\nn\\
V_{IV}^{(1)}= \dfrac{\gamma_1}{\sqrt{xy}} +\dfrac{\gamma_2(\sqrt{x}-\sqrt{y})}{\sqrt{xy}}+\dfrac{\gamma_3}{\sqrt{xy}(\sqrt{x}+\sqrt{y})^2}\,,
\qquad
&S=\left(
  \begin{smallmatrix}
    1 & q_2^2/4 \\
    1/q_1 & 1
  \end{smallmatrix}\right),\nn\\
V_{IV}^{(2)}= \gamma_1 xy+\gamma_2(x-y)+\dfrac{\gamma_3}{(x+y)^2}\,,
\qquad
&S=\left(
  \begin{smallmatrix}
    a/q_1 & b \\
    1/q_1 & 1
  \end{smallmatrix}\right)\nn
\end{array}
\]
for which integrals of motion $K_m$ and $K_\ell$ (\ref{k-m}-\ref{k-ell}) are  fifth and sixth order polynomials in the momenta.

Solution of the equations $K_m=\pm\{H_2,K_{m-1}\}$ and $\{H_1,K_{m-1}\}=0$ looks like
\ben
K_{m-1}&=&4p_1p_2u_1u_2(2fv_2+g_2)
+4v_2(2fv_1+g_1)(f v_2+g_2)+(4fh_2+g_2^2)v_1\nn\\
&=& 4\mu_2(\mu_1P'_2+P'_1)-(4fh_2-g_2^2)\lambda_1-4h_2g_1\,.\nn
\en
It is additional integral of motion, which is second order polynomial  for the system with potential  $V_{III}$ and fourth order polynomial  in the momenta for the other systems.

Now we present some superintegrable St\"ackel systems at $\kappa_1=1$ and $\kappa_2=3$. Here is one system with cubic additional integral $K_m$ (\ref{k-m})
\[
V_{III}^{(1)}= \gamma_1(x+2y)(2x+y)+\gamma_2(x+2y)+ \gamma_3(2x+y)\,,\qquad
S=\left(
  \begin{smallmatrix}
    a & b \\
    1 & 1
  \end{smallmatrix}
\right),
\]
and seven systems with the real potentials
\[\begin{array}{ll}
V_{I}=  \gamma_1(x+2y)(2x+y)+\dfrac{\gamma_2}{(x+y)^2}+\dfrac{\gamma_3}{(x-y)^2}\,,
\qquad
&S=\left(
  \begin{smallmatrix}
    a/q_1 & b/q_2 \\
    1/q_1 & 3/q_2
  \end{smallmatrix}
\right),
\nn\\
V_{II}^{(1)}= \gamma_1 xy+\dfrac{\gamma_2}{x^{2/3}y^{4/3}}+\dfrac{\gamma_3}{x^{1/3} y^{5/3}},
\qquad
&S=\left(
  \begin{smallmatrix}
      1/q_1 & 0 \\
    1/q_1^2 & 9/q_2^2
  \end{smallmatrix}
\right),\nn\\
V_{II}^{(2)}= \dfrac{\gamma_1}{\sqrt{xy}}+\dfrac{\gamma_2\sqrt{y}}{x^{5/2}}+\dfrac{\gamma_3 y^2} {x^{4}},
\qquad
&S=\left(
  \begin{smallmatrix}
    0 & 1 \\
    1/q_1^2 & 9/q_2^2
  \end{smallmatrix}\right)\nn\\
V_{II}^{(3)}= \dfrac{\gamma_1}{\sqrt{xy}}+\dfrac{\gamma_2}{x^{4/3}y^{2/3}}+\dfrac{\gamma_3} {x^{7/6}y^{5/6}},
\qquad
&S=\left(
  \begin{smallmatrix}
    1 & 0 \\
    1/q_1^2 & 9/q_2^2
  \end{smallmatrix}\right),\nn\\
V_{II}^{(4)}= \gamma_1 xy+\dfrac{\gamma_2 y^2}{x^4}+\dfrac{\gamma_3 y^5}{x^7},
\qquad
&S=\left(
  \begin{smallmatrix}
    0 & 1/q_2 \\
    1/q_1^2 & 9/q_2^2
  \end{smallmatrix}\right),\nn\\
V_{III}^{(2)}= \gamma_1(x^2-5x\sqrt{y\,}+4y)+\dfrac{\gamma_2 x}{\sqrt{y\,}}+\dfrac{\gamma_3}{\sqrt{y\,}}\,,\qquad
&S=\left(
  \begin{smallmatrix}
    2q_1 & 2q_2 \\
    1 & 1
  \end{smallmatrix}\right),\nn\\
V_{IV}=\gamma_1(x+5y)(5x+y)+\gamma_2(x-y)+\dfrac{\gamma_3}{(x-y)^2}\,,
\qquad
&S=\left(
  \begin{smallmatrix}
    a/q_1 & b \\
    1/q_1 & 1
  \end{smallmatrix}\right)\nn
\end{array}\]
for which integrals of motion $K_m$ and $K_\ell$ (\ref{k-m}-\ref{k-ell}) are  seventh and eights order polynomials in the momenta.

The algebra of integrals of motion $H_{1,2}$ and $K_m$ (\ref{k-m}) is the fifth or seventh order polynomial algebra in terms of the coefficients of the hyperelliptic curves
\bq
\{H_2,K_m\}=2K_\ell,\qquad \{H_2,K_\ell\}=2fK_m,\qquad
\{K_m,K_\ell\}=\pm F_Z\,,\nn
\eq
where polynomial $F_Z$ depends on the type of solution (\ref{uv-sol}) only:
\ben
F_{I}&=&2(4fh_2-g_2^2)^{\kappa_2-\kappa_1}\Bigl(4f(\kappa_1^2h_2g_1+\kappa_2^2h_1g_2)-g_1g_2(\kappa_2^2g_1+\kappa_1^2g_2)\Bigr)\,,
\nn\\
F_{II}&=&4(4fh_2-g_2^2)^{\kappa_2-\kappa_1}\Bigl(4f(\kappa_2+\kappa_1)h_2h_1-\kappa_1h_1g_2^2-\kappa_2h_2g_1^2\Bigr)\mp K_m^2\,,
\nn\\
F_{III}&=&4(4fh_2-g_2^2)^{\kappa_2-\kappa_1}\Bigl(4f(\kappa_1h_2+\kappa_2h_1)-\kappa_1g_2^2-\kappa_2g_1^2\Bigr)f\,,
\label{alg-comm}\\
F_{IV}&=&2(4fh_2-g_2^2)^{\kappa_2-\kappa_1}\Bigl(4f(2\kappa_2fh_1-\kappa_1h_2g_1)-2\kappa_2fg_1^2+\kappa_1g_1g_2^2\Bigr).
\nn
\en
Here choice of sign $+$ or $-$ depends on $\kappa$'s.

As above, the St\"ackel transformations (\ref{t-change}) relate systems associated with  one type of the solutions (\ref{uv-sol}), whereas algebra of integrals of motion is invariant with respect to such transformations.

The complete classification of such superintegrable systems requires further investigations.

\section{Conclusion}
We discuss an application of the addition theorem to construction of algebraic integrals of motion  from the multi-valued action-angle variables.

We propose new  algorithm to construction of the superintegrable St\"ackel systems associated with zero-genus hyperelliptic curves. It allows us to prove that the Drach classification of the St\"ackel systems with cubic integral of motion   (\ref{k3-dr}) associated with the addition theorem (\ref{add-ln}) is complete. Moreover, we present some new two-dimensional superintegrable systems with third, fifth and seventh order integrals of motion.

The proposed method may be applied to construction of the higher order additional polynomial integrals of motion for the $n$-dimensional superintegrable St\"ackel systems on the different manifolds. On the other hand, we prove that there are some superintegrable systems, which miss out of this construction. It will be interesting to study a mathematical mechanism of the appearance such superintegrable systems.

The research was partially supported by the RFBR grant 06-01-00140.

\end{document}